\documentclass[twocolumn]{aastex62}
\usepackage{graphicx}
\usepackage{amsmath}
\usepackage{color}

\newcommand{\cotwo}{\mbox{\rm CO(2-1)}} 
\newcommand{\hi}{\mbox{\rm H$\,$\scshape{i}}} 

\newcommand{\kmpers}{\mbox{km~s$^{-1}$}}

\newcommand{\acounits}{\mbox{\rm M$_{\odot}$ pc$^{-2}$} \mbox{(K km s$^{-1}$)$^{-1}$}}

\newcommand{\aco}{\mbox{$\alpha_{\rm CO}$}}
\newcommand{\sigv}{\mbox{$\sigma_{\rm v}$}}
\newcommand{\avir}{\mbox{$\alpha_{\rm vir}$}}

\newcommand{\Mmol}{\mbox{$M_{\rm mol}$}}

\newcommand{\Mmolsd}{\mbox{$\Sigma_{\rm mol}$}}

\newcommand{\Msun}{\mbox{$\rm M_{\odot}$}}
\newcommand{\Msunperpc}{\mbox{\rm M$_{\odot}$ pc$^{-2}$}}

\newcommand{\SFR}{\mbox{\rm SFR}}
\newcommand{\SFRsd}{\mbox{$\Sigma_{\rm SFR}$}}
\newcommand{\Msunperyr}{\mbox{\rm M$_{\odot}$ yr$^{-1}$}}

\newcommand{\tdep}{\mbox{$\tau_{\rm dep}$}}
\newcommand{\tdyn}{\mbox{$\tau_{\rm dyn}$}}
\newcommand{\tcross}{\mbox{$\tau_{\rm cross}$}}
\newcommand{\tff}{\mbox{$\tau_{\rm ff}$}}
\newcommand{\edyn}{\mbox{$\epsilon_{\rm dyn}$}}
\newcommand{\eff}{\mbox{$\epsilon_{\rm ff}$}}

\accepted{2019 August 8}
\submitjournal{The Astrophysical Journal}

\begin{document}

\title{How Galactic Environment affects the Dynamical State of Molecular Clouds and their Star Formation Efficiency}

\shorttitle{Galactic Environment and Star Formation}
\shortauthors{Schruba, Kruijssen, Leroy}

\author{Andreas Schruba}
\affiliation{Max-Planck-Institut f\"ur extraterrestrische Physik, Giessenbachstra{\ss}e 1, D-85748 Garching, Germany}

\author{J.~M.\ Diederik Kruijssen}
\affiliation{
Astronomisches Rechen-Institut, Zentrum f\"{u}r Astronomie der Universit\"{a}t Heidelberg, M\"{o}nchhofstra{\ss}e 12-14, D-69120 Heidelberg, Germany 
}
\affiliation{Max-Planck-Institut f\"ur Astronomie, K\"{o}nigstuhl 17, D-69117 Heidelberg, Germany}

\author{Adam K.\ Leroy}
\affiliation{Department of Astronomy, The Ohio State University, 140 W 18th St, Columbus, OH 43210, USA}

\correspondingauthor{Andreas Schruba}
\email{schruba@mpe.mpg.de}

\begin{abstract} 
We investigate how the dynamical state of molecular clouds relates to host galaxy environment, and how this impacts the star formation efficiency in the Milky Way and seven nearby galaxies. We compile measurements of molecular cloud and host galaxy properties and determine mass-weighted mean cloud properties for entire galaxies and distinct subregions within. We find molecular clouds to be in ambient pressure-balanced virial equilibrium, where clouds in gas-rich, molecular-dominated, high-pressure regions are close to self-virialization, whereas clouds in gas-poor, atomic-dominated, low-pressure environments achieve a balance between their internal kinetic pressure and external pressure from the ambient medium. The star formation efficiency per free-fall time of molecular clouds is low ${\sim}0.1\%{-}1\%$ and shows systematic variations of $2$~dex as a function of the virial parameter and host galactic environment. The trend observed for clouds in low-pressure environments---as the solar neighborhood---is well matched by state-of-the-art turbulence-regulated models of star formation. However, these models substantially overpredict the low observed star formation efficiencies of clouds in high-pressure environments, which suggests the importance of additional physical parameters not yet considered by these models.
\end{abstract}


\section{Introduction}
\label{sec:intro}

Stars form in molecular clouds. For galaxies on the star formation main sequence \citep[see reviews by][]{Blanton09, Renzini15}, the molecular gas mass (\Mmol) and star formation rate (\SFR) are tightly correlated \citep[see reviews by][]{Kennicutt12, Krumholz14}. This tight correlation holds for entire galaxies \citep{Kennicutt98b, Young95, Saintonge11a} and down to kiloparsec-scale regions \citep{Bigiel11, Leroy13b}, below which differences in the evolutionary state of individual star-forming regions introduces significant scatter \citep{Schruba10, Onodera10, Feldmann11b, Kruijssen14a, Kreckel18}.

For main sequence galaxies, the molecular gas depletion time, $\tdep = \Mmolsd / \SFRsd$, is of order $1{-}2$~Gyr \citep{Leroy13b} and has experienced only modest evolution through cosmic times \citep{Saintonge13}. This implies that star formation on galactic scales is highly inefficient, with only a few per cent of a galaxy's gas mass being converted to stars per free-fall time over the disk scale height \citep{Krumholz12}. Importantly, \tdep\ shows systematic second-order variations: longer \tdep\ are found in massive, bulge-dominated spiral galaxies \citep{Saintonge11b, Shi11}, early-type galaxies \citep{Wei10, Davis14} and (potentially) low-mass dwarf galaxies (\citealp{Hunt15, Amorin16, Grossi16}; but see \citealp{Filho16}) while shorter \tdep\ are found in (many) galaxy centers \citep{Leroy13b} and gas-rich early-universe galaxies \citep{Tacconi13, Tacconi18}. Understanding what regulates the \mbox{(in-)}efficiency of star formation and results in the observed gas--SFR relationship, its normalization (i.e.,~\tdep), and variations thereof in different galactic environments is therefore a key task to understand galaxy evolution.

\emph{Disk equilibrium models} try to explain a galaxy's SFR by a dynamical balance of gravitational forces promoting star formation and energy and momentum feedback by recently formed stars counteracting gravitational collapse \citep{Toomre64, Ostriker10, Kim11, Hopkins11, Romeo11, FaucherGiguere13, Agertz15b, Hayward17, Krumholz17}. In these models, the SFR self-regulates to produce stellar feedback just sufficient to keep the galaxy's gas disk in vertical pressure and energy balance and marginally stable against radial instabilities. As the radial and vertical distribution of stars, gas, and dark matter in the galaxy vary, so does the SFR per unit gas mass needed to maintain equilibrium. This leads to different values of \tdep\ in galaxy centers, outer disks, and dwarf galaxies (see above references).

Limitations of these models have been the dependence on phenomenological star formation prescriptions and a highly uncertain feedback efficiency. They have also tended to adopt  a simplified modeling of the ISM structure relevant to gravitational instability \citep{Romeo11, Agertz15b} and treat the cold, star-forming gas in a simple way. To date, the fraction of cold molecular, star-forming gas has been represented either as a pressure-regulated \citep{Ostriker10, Kim11} or shielding-regulated two-phase medium \citep{Krumholz13}, with little distinction between different physical states within the molecular gas. Recently, the first models have been presented that also consider the dynamical state (i.e., gravitational boundedness) of the cold, star-forming gas in a galactic context \citep{Semenov16, Krumholz18}. However, a fully self-consistent theory of galactic and cloud-scale properties with essentially no meaningful free parameters remains to be developed.

\emph{Turbulent cloud models} attempt to predict the galaxy's SFR from the star formation efficiency (SFE)\footnote{In turbulent models, the SFE is typically defined as the fraction of molecular gas mass converted to stars over some fiducial timescale, which is usually the gravitational free-fall time.} of individual turbulent molecular clouds \citep[Eq.~\ref{eq:sfeq};][]{Krumholz05, Padoan11, Hennebelle11, Federrath12}. These represent the smaller-scale complement to the disk equilibrium models. In these models the mean density, gravitational boundedness, and Mach number play a central role as they define the cloud's density structure, balance of kinetic and gravitational energy, and the mass in self-gravitating dense clumps where stars form. Star formation in these clouds progresses on a few dynamical timescales, typically parameterized by the gravitational free-fall time ($\tff \propto \rho^{-0.5}$), though at a low efficiency per free-fall time, $\eff \approx 1\%$ \citep{Krumholz07}.

Observable properties of molecular clouds, such as their size ($R$), velocity dispersion (\sigv), and surface density ($\Sigma$), offer a snapshot view of their dynamical state, which the models employ to predict the cloud's SFE. Early observational work in the Milky Way suggested a common set of cloud properties described by the size--line width relation, virial equilibrium, and constant surface density \citep{Larson81, Solomon87}. Though supported by early extragalactic observations \citep{Rosolowsky03, Bolatto08}, subsequent observations revealed substantial variations in the properties of molecular clouds in the Galactic center \citep{Oka01, Shetty12, Kruijssen13, Walker16, Kauffmann16a}, the inner and outer Milky Way \citep{Heyer09, Rice16, MivilleDeschenes17}, and across nearby star-forming galaxies \citep{Hughes13b, DonovanMeyer13, Leroy13a, Colombo14a, Leroy15a, Leroy16, Sun18}. Meanwhile, starburst galaxies and merging systems have long been observed to show high surface and volume densities \citep[e.g.,][among many others]{Downes98, Wilson03}. Today it is clear that there is no common set of cloud properties but that these systematically vary with galactic properties. We are still in the early stages of understanding how these variations in cloud properties affect a cloud's SFE.

The turbulence regulated models of star formation have been designed to match the SFRs found in (idealized) numerical simulations of ISM turbulence \citep[][and many others]{Padoan11, Padoan12, Federrath13c}. Their validation by observations remains inconclusive, however. The models agree with observations of (mostly low-mass) star-forming regions in the solar neighborhood \citep[see compilations by][]{Federrath12, Hennebelle13, Krumholz14, Padoan14}. Subsequent studies of larger giant molecular cloud (GMC) samples in the Milky Way and the LMC revealed inconsistencies between the absolute SFR predicted by these models and observations. These studies also found significant scatter in the SFE of individual clouds, far in excess to what the models predict (these have been attributed to an accelerating SFR along a cloud's time evolution), and only a weak dependence of the SFE on cloud properties \citep{Murray11, Evans14, LeeE16, Vutisalchavakul16, Ochsendorf17}. Comparing cloud population averages to the SFR-per-H$_2$ and \eff\ in M51, \citet{Leroy17b} also found a poor match to the predictions of turbulent models. The sense of the correlations between cloud properties and \eff\ in M51 appears opposite that present in some of the models. They also found lower absolute values of \eff\ than either the models or Galactic observations. A recent study of star formation in Galactic center clouds finds the turbulence regulated models to agree with observations but they also stress that the attempt of falsifying the models is obstructed by the lack of consensus on the values of their free parameters \citep{Barnes17}.

The above results demonstrate that we are still in the early stages of confronting turbulent models with observations. And the link between the disk equilibrium models and the cloud properties relevant to the turbulent models remains even less well understood. Extensive observational and theoretical work is required to establish the link between the galactic SFR, disk structure, cloud-scale gas properties, and star formation on the scale of individual clouds. In the coming years, we expect that uniform surveys of diverse environments with ALMA will play a major role in such experiments. For example, the ongoing PHANGS-ALMA survey\footnote{\label{footnote2}For information on PHANGS, see \href{http://phangs.org}{\tt http://phangs.org}} (``Physics at High Angular-resolution in Nearby GalaxieS with ALMA''; A.~K.\ Leroy et al., in preparation) is mapping molecular gas via \cotwo\ emission at cloud scales in a large sample of $74$ nearby star-forming galaxies. This allows for a uniform statistical analysis of ISM and cloud properties and their connection to star formation \citep[][and work in preparation by M.~Chevance et~al.; E.~Rosolowsky et~al.; E.~Schinnerer et~al.; A.~Schruba et~al.; J.~Sun et~al.; D.~Utomo et~al.]{Sun18, Utomo18}.

At the moment, however, a valuable, largely untapped resource exists in the form of recent large surveys of molecular clouds in the Milky Way and the nearest galaxies. In this paper, we synthesize the current best-in-class single galaxy GMC studies, estimate the environments hosting the cloud populations, and compare them to theoretical expectations for the dynamical state and star formation efficiency of the gas. We present a literature compilation of galactic disk and molecular cloud properties for the Milky Way and seven local galaxies. We use these to address two main questions:

{\em How does the dynamical state of the gas depend on environment?} We compare the cloud's dynamical state (i.e., the observed virial parameter) to its expected value if the cloud is in pressure equilibrium with its local galactic environment. Similar comparisons have been performed, e.g., for dense clumps inside Galactic molecular clouds \citep{Bertoldi92}, for whole molecular clouds in the Milky Way \citep{Field11}, for the LMC, M33, and M51 \citep{Hughes13b, Hughes16}. Here we extend this work to a sample of eight local galaxies broken into several discrete environments.

{\em How does the apparent star formation efficiency per free-fall time ($\eff$) relate to the properties of the local cloud population?} We relate mass-weighted average cloud properties to the gas depletion time ($M_{\rm mol}/{\rm SFR}$) for each of our targets. We contrast these to the predictions of a suite of turbulence regulated models of star formation as summarized in \citet{Federrath12}.

This synthesis should represent the most complete, most direct link between environment and cloud dynamical state, and the most general test of turbulent models of star formation to date.

We start by reviewing the virial theorem that defines the cloud's dynamical state (Section~\ref{sec:virialtheorem}) and the turbulent star formation models (Section~\ref{sec:theories}). Then we present our literature compilation of galactic and cloud properties (Section~\ref{sec:data}, Table~\ref{t1}), link the cloud's dynamical state to galactic environment (Section~\ref{sec:dynstate}), and compare the SFR predicted by the turbulence regulated models to the observed SFR (Sections~\ref{sec:sfedyn} and~\ref{sec:comparison}). We conclude by giving an outlook on future work (Section~\ref{sec:outlook}) and summarizing our findings (Section~\ref{sec:summary}).

\subsection{Virial Theorem for Molecular Clouds}
\label{sec:virialtheorem}

The energy balance and thus dynamical state of a molecular cloud is described by the virial theorem. For a non-magnetized, isothermal, self-gravitating spherical cloud immersed in a uniform external medium, the virial theorem is \citep[e.g.,][]{Spitzer78}
\begin{equation}
\frac{1}{2} \frac{{\rm d}^2 I}{{\rm d}t^2} \enskip
= \enskip 2(\mathcal{T} - \mathcal{T}_S) + \mathcal{W}~.
\end{equation}
\noindent Here $I$ is the moment of inertia within the cloud's volume. The kinetic energy of the cloud is $2\mathcal{T} = 3M\sigma_\mathrm{v}^2$, with $\sigma_{\rm v}$ being the one-dimensional velocity dispersion of the gas. The surface term of the kinetic energy is $2\mathcal{T}_S = 4\pi R^3 P_\mathrm{ext}$, where $P_{\rm ext}$ refers to the ambient gas pressure. The self-gravitational binding energy of the cloud is $\mathcal{W} = - \Gamma G M^2 / R$, where $\Gamma$ is a geometrical form factor. $\Gamma = 0.6$ for a constant density sphere (assumed here) and $0.73$ for an isothermal sphere of maximal critical mass \citep[][]{Elmegreen89}. The gravitational constant in our units is $G = 1 / 232.5\ \Msun^{-1}\ {\rm pc}\ {\rm km}^2\ {\rm s}^{-2}$.

The sign of ${\rm d}^2 I / {\rm d}t^2$ determines whether the cloud will contract or expand, so that the balance of the terms on the right-hand side defines its imminent dynamical evolution. Following \citet[][and many others]{Bertoldi92}, we use the dimensionless viral parameter
\begin{equation}\label{eq:avirobs}
\alpha_{\rm vir,\,obs}  = \frac{2 \mathcal{T}}{\mathcal{|W|}} = \frac{3 \sigma^2}{\pi \Gamma G R \Sigma} = 
\frac{P_\mathrm{int}}{P_\mathrm{self}}
\end{equation}
\noindent to describe the relative importance of the cloud's kinetic energy and gravitational energy. In a complementary way, we express the theoretically expected virial parameter for a cloud in pressure equilibrium with its ambient medium by
\begin{equation}\label{eq:avirtheo}
\alpha_{\rm vir,\,theo} = 1 + \frac{2 \mathcal{T}_S}{\mathcal{|W|}} = 1 + \frac{P_\mathrm{ext}}{P_\mathrm{self}}~.
\end{equation}
\noindent Here we have also defined the cloud's internal kinetic pressure
\begin{equation}\label{eq:Pint}
P_{\rm int} = 3 \rho \sigma_{\rm v}^2
\end{equation}
\noindent and the pressure related to the cloud's self-gravitational binding energy
\begin{equation}\label{eq:Pself}
P_\mathrm{self} = \frac{\Gamma}{2} \frac{\pi G}{2} \Sigma^2~.
\end{equation}
\noindent Here $\Sigma$ refers to the mass surface density of the cloud.

We estimate the ambient midplane gas pressure from the weight of the gas in the combined gaseous and stellar potential, assuming vertical hydrostatic equilibrium \citep[][see also \citealp{Blitz04,Ostriker10,Field11, Kim13}]{Elmegreen89}
\begin{equation}\label{eq:Pext}
P_\mathrm{ext} \enskip = \enskip \frac{\pi \mathrm{G}}{2} \Sigma_\mathrm{ism}^2
\left( 1 + \frac{\sigma_\mathrm{ism}}{\sigma_\star} \frac{\Sigma_\star}{\Sigma_\mathrm{ism}} \right)~.
\end{equation}
Here $\Sigma_{\rm ism}$ and $\Sigma_\star$ are the average surface densities of the diffuse interstellar medium and stars, and the ratio $\sigma_{\rm ism} / \sigma_\star$ reduces the gravitational force exerted by stars onto the gas according to the differences in gaseous and stellar scale heights.

Following \citet{Ostriker10} and \citet{Kim13} we assume that the diffuse ISM provides the pressure onto molecular clouds. We calculate the average surface density of the diffuse ISM as

\begin{equation}\label{eq:sdism}
\Sigma_{\rm ism} = \Sigma_{\rm atom} + {(1-f_{\rm gmc})} \Sigma_{\rm mol}~,
\end{equation}
where $f_{\rm gmc}$ is the fraction of molecular gas in the adopted GMC catalog (Table~\ref{t1}). This treatment assumes that all atomic gas is diffuse and that all GMCs are self-gravitating and so not part of the diffuse medium. Our treatment neglects the likely minor contribution of bound GMCs to the galactic disk potential \citep[see][for a complete treatment]{Ostriker10}. We also assume that the contributions of cosmic rays and magnetic fields play a minor role in supporting the gaseous disk \citep[cf.][]{Elmegreen89}.

\defcitealias{Hennebelle11}{HC11}
\defcitealias{Krumholz05}{KM05}
\defcitealias{Padoan11}{PN11}
\defcitealias{Padoan12}{P12}
\defcitealias{Federrath12}{FK12}

\subsection{Theories of Star Formation in Molecular Clouds}
\label{sec:theories}

The turbulence regulated star formation models predict the SFR in a galaxy or part of a galaxy by scaling the ratio of cloud mass, $M_{\rm gmc}$, and free-fall time, $\tau_{\rm ff}$ by the \emph{star formation efficiency per free-fall time},~$\epsilon_{\rm ff}$,
\begin{equation}\label{eq:sfeq}
{\rm SFR} = \epsilon_{\rm ff} \frac{M_{\rm gmc}}{\tau_{\rm ff} (\rho)}~.
\end{equation}
Here the free-fall time is given by
\begin{equation}\label{eq:tff}
\tau_{\rm ff}(\rho) = {\left( \frac{3 \pi}{32 G \rho} \right)}^{1/2}~,
\end{equation}
and is evaluated either at the mean cloud density or in a multi-scale fashion at the volume density of each gas parcel inside the cloud.

We consider the models by \citet[][hereafter \citetalias{Krumholz05}]{Krumholz05}, \citet[][\citetalias{Padoan11}]{Padoan11}, and \citet[][\citetalias{Hennebelle11}]{Hennebelle11} in the form\footnote{\label{footnote3}We note that the \citeauthor{Krumholz05} model differs in \citetalias{Krumholz05} and \citet[][\citetalias{Federrath12}]{Federrath12} by the definition of the sonic length scale $\lambda_{\rm s}$: the former use the one-dimensional velocity dispersion while the latter use the three-dimensional one. Here we adopt the \citetalias{Krumholz05} definition implying SFEs per free-fall time that are a factor ${\sim}10$ larger than for the \citetalias{Federrath12} definition. We note that \citet{Barnes17} adopt the \citetalias{Federrath12} definition.} presented by \citet[][their table~1; see also \citealt{Padoan14}]{Federrath12}, as well as the simplified empirical fit to the \citetalias{Padoan11} model by \citet[][\citetalias{Padoan12}]{Padoan12}.

These models derive \eff\ by integrating the cloud's density distribution to obtain the mass fraction above some critical density for collapse. That mass fraction is then compared to the free-fall time to generate a rate of star formation. The models differ on how the density distribution and the critical density for collapse depend on the mean physical properties of a cloud, i.e., the virial parameter ($\alpha_{\rm vir}$), the sonic Mach number\footnote{In this paper we adopt three-dimensional Mach numbers.} ($\mathcal{M}$), the turbulence driving parameter ($b$), the relative strength of thermal and magnetic pressure ($\beta$). They also differ regarding whether the free-fall time is calculated at a fixed density (single free-fall time models) or at the local density along the density distribution (multi free-fall time models).

To compare the models with observations, we need to choose values of their free parameters. Here we adopt the fiducial parameters from the original works: $\epsilon_{\rm core} = 0.5$, $\phi_t = 1.91$, $\phi_x = 1.12$, $\theta = 0.35$, and $y_{\rm cut} = 0.1$. Note that \citet{Federrath12} derived significantly different values for these parameters using magnetohydrodynamical turbulent box simulations. We adopt the following default cloud properties: $\mathcal{M} = 10$, typical for modestly supersonic molecular gas \citep{Padoan14} and Milky Way GMCs \citep{Heyer09}, $b = 0.4$ found for a mix of solenoidal and compressive forcing \citep{Federrath10a}, and negligible magnetic fields, $\beta \rightarrow \infty$ \citep{Crutcher12}. We will also consider the effect of varying these quantities about their default values. We treat the virial ratio as known, determined by observations.

\section{Data}
\label{sec:data}

Table~\ref{t1} presents our compilation of galaxy- and cloud-scale measurements. Together these allow us to assess (a) the relationship between the cloud's dynamical state and local galactic environment and (b) SFE per free-fall time as a function of the local cloud population.

As the table shows,  surveys of the molecular gas with resolution to study individual molecular clouds now cover a suite of individual nearby galaxies. Compared to the first studies targeting the local group \citep[see review by][]{Fukui10} these have improved resolution and sensitivity and span a larger range of environments within galaxies. We combine surveys of the LMC \citep[using MOPRA;][]{Wong09}, M33 \citep[using the IRAM \mbox{30-m};][]{Gratier10b, Druard14}, M31 \citep[using CARMA;][]{CalduPrimo16,Schruba19b}, the lenticular galaxy NGC4526 \citep{Utomo15}, M51 \citep[using NOEMA;][]{Schinnerer13, Pety13}, the central starburst in NGC253 \citep{Leroy15a}, and the nearby dwarf spiral NGC300 \citep[][]{Kruijssen19b, Schruba19b}.

\movetabledown=2.8in
\setlength{\tabcolsep}{3pt}
\renewcommand{\arraystretch}{1.1}
\begin{deluxetable*}{ll*{12}{R@{\hskip 2pt}l}}
\rotate
\vspace*{-0.3in}

\tablewidth{0pt}
\tabletypesize{\scriptsize}
\tablecolumns{26}
\tablecaption{Galaxy and Cloud Properties\label{t1}}
\tablehead{\colhead{Parameter} & \colhead{Unit} & \twocolhead{MW\_CMZ} & \twocolhead{MW\_5kpc} & \twocolhead{MW\_10kpc} & \twocolhead{LMC} & \twocolhead{M33} & \twocolhead{M31\_6kpc} & \twocolhead{M31\_11kpc} & \twocolhead{NGC300} & \twocolhead{M51\_1kpc} & \twocolhead{M51\_3kpc} & \twocolhead{NGC253} & \twocolhead{NGC4526}}
\startdata
\decimals
\hfill\\[-5px]
\cutinhead{\footnotesize Global Galaxy Properties}
Dist & Mpc &  0.008 & (\citenum{BlandHawthorn16}) &  0.005 & (\citenum{Rice16}) &  0.005 & (\citenum{Rice16}) &  0.05 & (\citenum{Jameson16}) &  0.84 & (\citenum{Kam15}) &  0.78 & (\citenum{Dalcanton12}) &  0.78 & (\citenum{Dalcanton12}) &  1.90 & (\citenum{Gieren04}) &  7.60 & (\citenum{Ciardullo02}) &  7.60 & (\citenum{Ciardullo02}) &  3.50 & (\citenum{Rekola05}) & 16.40 & (\citenum{Tonry01}) \\
log M$_\star$ & \Msun &  9.15 & (\citenum{Launhardt02}) & 10.76 & (\citenum{Licquia16}) & 10.76 & (\citenum{Licquia16}) &  9.30 & (\citenum{Skibba12}) &  9.70 & (\citenum{Seigar11}) & 11.00 & (\citenum{Tamm12}) & 11.00 & (\citenum{Tamm12}) &  9.30 & (\citenum{MunozMateos07}) & 10.56 & (\citenum{Leroy08}) & 10.56 & (\citenum{Leroy08}) & 10.33 & (\citenum{Lucero15}) & 10.97 & (\citenum{Amblard14}) \\
log M$_\mathrm{atom}$ & \Msun & \nodata &  &  9.90 & (\citenum{Kalberla09}) &  9.90 & (\citenum{Kalberla09}) &  8.80 & (\citenum{StaveleySmith03}) &  9.10 & (\citenum{Gratier10b}) &  9.90 & (\citenum{Braun09}) &  9.90 & (\citenum{Braun09}) &  9.30 & (\citenum{Westmeier11}) &  9.45 & (\citenum{Leroy08}) &  9.45 & (\citenum{Leroy08}) &  9.45 & (\citenum{Lucero15}) & \nodata &  \\
log M$_\mathrm{mol}$ & \Msun &  7.30 & (\citenum{Longmore13}) &  9.17 & (\citenum{Heyer15}) &  9.17 & (\citenum{Heyer15}) &  7.80 & (\citenum{Jameson16}) &  8.50 & (\citenum{Druard14}) &  8.80 & (\citenum{Nieten06}) &  8.80 & (\citenum{Nieten06}) &  8.10 & (\citenum{Schruba19b}) &  9.82 & (\citenum{Schruba12}) &  9.82 & (\citenum{Schruba12}) &  8.55 & (\citenum{Leroy15a}) &  8.59 & (\citenum{Young11}) \\
log SFR & \Msunperyr & -1.05 & (\citenum{Barnes17}) &  0.22 & (\citenum{Licquia15}) &  0.22 & (\citenum{Licquia15}) & -0.70 & (\citenum{Jameson16}) & -0.35 & (\citenum{Verley09}) & -0.60 & (\citenum{Ford13}) & -0.60 & (\citenum{Ford13}) & -0.72 & (\citenum{Kang16}) &  0.47 & (\citenum{Schruba12}) &  0.47 & (\citenum{Schruba12}) &  0.62 & (\citenum{Sanders03}) & -0.76 & (\citenum{Amblard14}) \\
\cutinhead{\footnotesize Local Galaxy Properties}
R$_\mathrm{gal}$ & kpc & 0{-}0.12 & (\citenum{Longmore13}) & 2{-}8 & (\citenum{Rice16}) & 9{-}15 & (\citenum{Rice16}) & 0{-}4 & (\citenum{Wong09}) & 0{-}7 & (\citenum{Druard14}) & 5{-}8 & (\citenum{Schruba19b}) & 9{-}14 & (\citenum{Schruba19b}) & 0{-}4 & (\citenum{Schruba19b}) & 0{-}1.5 & (\citenum{Colombo14a}) & 1.5{-}5 & (\citenum{Colombo14a}) & 0{-}0.3 & (\citenum{Leroy15a}) & 0{-}1 & (\citenum{Utomo15}) \\
$\sigma_\mathrm{v,star}$ & \kmpers & 100.0 & (\citenum{Launhardt02}) &  35.0 & (\citenum{Bovy13}) &  15.0 & (\citenum{Bovy13}) &  20.0 & (\citenum{vanderMarel02}) &  21.0 & (\citenum{Hughes13b}) &  70.0 & (\citenum{Dorman15}) &  40.0 & (\citenum{Dorman15}) &  20.0 & (\citenum{Schruba19a}) &  96.0 & (\citenum{Hughes13b}) &  70.0 & (\citenum{Schruba19a}) &  30.0 & (\citenum{Leroy15a}) & 233.0 & (\citenum{Utomo15}) \\
$\sigma_\mathrm{v,gas}$ & \kmpers &  15.0 & (\citenum{Kruijssen14b}) &  15.0 & (\citenum{Schruba19a}) &  15.0 & (\citenum{Schruba19a}) &  10.0 & (\citenum{Wong09}) &  15.0 & (\citenum{Schruba19a}) &  17.0 & (\citenum{Braun09}) &  17.0 & (\citenum{Braun09}) &  15.0 & (\citenum{Schruba19a}) &  20.0 & (\citenum{Schuster07}) &  16.0 & (\citenum{CalduPrimo13}) &  30.0 & (\citenum{Leroy15a}) &  10.0 & (\citenum{Utomo15}) \\
$\Sigma_\star$ & \Msunperpc & 3800.0 & (\citenum{Kruijssen14b}) &  250.0 & (\citenum{BlandHawthorn16}) &   25.0 & (\citenum{BlandHawthorn16}) &   46.0 & (\citenum{Jameson16}) &   85.0 & (\citenum{Seigar11}) &  170.0 & (\citenum{Schruba19b}) &   50.0 & (\citenum{Schruba19b}) &   26.0 & (\citenum{MunozMateos15}) & 1000.0 & (\citenum{Pety13}) &  200.0 & (\citenum{Pety13}) & 2750.0 & (\citenum{Iodice14}) &  800.0 & (\citenum{Utomo15}) \\
$\Sigma_\mathrm{atom}$ & \Msunperpc & \nodata &  &   10.0 & (\citenum{Kalberla09}) &   16.0 & (\citenum{Nakanishi16}) &   18.0 & (\citenum{StaveleySmith03}) &    7.0 & (\citenum{Druard14}) &    3.4 & (\citenum{Schruba19b}) &   10.0 & (\citenum{Schruba19b}) &    7.0 & (\citenum{Westmeier11}) &    6.0 & (\citenum{Schruba19b}) &    9.0 & (\citenum{Schruba19b}) &    3.0 & (\citenum{Lucero15}) & \nodata &  \\
$\Sigma_\mathrm{mol}$ & \Msunperpc & 1000.0 & (\citenum{Henshaw16}) &    6.5 & (\citenum{Heyer15}) &    3.0 & (\citenum{Sofue16}) &    2.0 & (\citenum{Jameson16}) &    3.0 & (\citenum{Druard14}) &    1.7 & (\citenum{Schruba19b}) &    1.4 & (\citenum{Schruba19b}) &    2.2 & (\citenum{Schruba19b}) &  175.0 & (\citenum{Pety13}) &   60.0 & (\citenum{Pety13}) &  700.0 & (\citenum{Leroy15a}) &  175.0 & (\citenum{Davis14}) \\
$\Sigma_\mathrm{sfr}$ & \Msunperpc\ Gyr$^{-1}$ & 4500.0 & (\citenum{Barnes17}) &    4.5 & (\citenum{Kennicutt12}) &    1.5 & (\citenum{Kennicutt12}) &    5.0 & (\citenum{Jameson16}) &    3.0 & (\citenum{Schruba10}) &    1.6 & (\citenum{Schruba19b}) &    3.3 & (\citenum{Schruba19b}) &    3.1 & (\citenum{Schruba19b}) &  115.0 & (\citenum{Meidt13}) &   24.0 & (\citenum{Meidt13}) & 3500.0 & (\citenum{Leroy15a}) &   90.0 & (\citenum{Davis14}) \\
$\tau_\mathrm{dep}$ & Gyr &  0.22 & (\citenum{Barnes17}) &  0.90 & (\citenum{Kennicutt12}) &  0.90 & (\citenum{Kennicutt12}) &  0.32 & (\citenum{Jameson16}) &  1.00 & (\citenum{Schruba10}) &  4.40 & (\citenum{Schruba19b}) &  1.95 & (\citenum{Schruba19b}) &  0.71 & (\citenum{Schruba19b}) &  1.50 & (\citenum{Meidt13}) &  2.50 & (\citenum{Meidt13}) &  0.20 & (\citenum{Leroy15a}) &  1.90 & (\citenum{Davis14}) \\
\cutinhead{\footnotesize Local Cloud Properties}
F$_\mathrm{gmc}$ & &  0.10 & (\citenum{Henshaw16}) &  0.35 & (\citenum{Rice16}) &  0.20 & (\citenum{Rice16}) &  0.30 & (\citenum{Wong11}) &  0.50 & (\citenum{Druard14}) &  0.34 & (\citenum{Schruba19a}) &  0.34 & (\citenum{Schruba19a}) &  0.70 & (\citenum{Schruba19a}) &  0.58 & (\citenum{Colombo14a}) &  0.56 & (\citenum{Colombo14a}) &  0.60 & (\citenum{Leroy15a}) &  0.24 & (\citenum{Utomo15}) \\
R & pc &   2.2 & (\citenum{Henshaw16}) &  78.3 & (\citenum{Rice16}) &  80.2 & (\citenum{Rice16}) &  28.6 & (\citenum{Wong11}) &  71.0 & (\citenum{Druard14}) &  45.9 & (\citenum{Schruba19a}) &  41.0 & (\citenum{Schruba19a}) &  38.0 & (\citenum{Schruba19a}) &  66.2 & (\citenum{Colombo14a}) &  67.7 & (\citenum{Colombo14a}) &  32.9 & (\citenum{Leroy15a}) &  21.5 & (\citenum{Utomo15}) \\
$\sigma_\mathrm{v}$ & \kmpers &  5.87 & (\citenum{Henshaw16}) &  4.73 & (\citenum{Rice16}) &  4.00 & (\citenum{Rice16}) &  2.69 & (\citenum{Wong11}) &  3.80 & (\citenum{Druard14}) &  5.20 & (\citenum{Schruba19a}) &  3.23 & (\citenum{Schruba19a}) &  3.43 & (\citenum{Schruba19a}) &  8.77 & (\citenum{Colombo14a}) &  7.89 & (\citenum{Colombo14a}) & 23.17 & (\citenum{Leroy15a}) &  8.94 & (\citenum{Utomo15}) \\
M$_\mathrm{lum}$ & $10^6$ \Msun &  0.01 & (\citenum{Henshaw16}) &  3.21 & (\citenum{Rice16}) &  0.91 & (\citenum{Rice16}) &  0.14 & (\citenum{Wong11}) &  0.52 & (\citenum{Druard14}) &  0.20 & (\citenum{Schruba19a}) &  0.15 & (\citenum{Schruba19a}) &  0.21 & (\citenum{Schruba19a}) &  5.46 & (\citenum{Colombo14a}) &  4.16 & (\citenum{Colombo14a}) & 34.60 & (\citenum{Leroy15a}) &  1.89 & (\citenum{Utomo15}) \\
M$_\mathrm{vir}$ & $10^6$ \Msun &  0.09 & (\citenum{Henshaw16}) &  2.34 & (\citenum{Rice16}) &  2.46 & (\citenum{Rice16}) &  0.24 & (\citenum{Wong11}) &  1.32 & (\citenum{Druard14}) &  1.53 & (\citenum{Schruba19a}) &  0.49 & (\citenum{Schruba19a}) &  0.51 & (\citenum{Schruba19a}) &  6.48 & (\citenum{Colombo14a}) &  5.52 & (\citenum{Colombo14a}) & 18.50 & (\citenum{Leroy15a}) &  1.81 & (\citenum{Utomo15}) \\
$\alpha_\mathrm{vir}$ & &  7.83 & (\citenum{Henshaw16}) &  1.29 & (\citenum{Rice16}) &  3.49 & (\citenum{Rice16}) &  3.17 & (\citenum{Wong11}) &  4.44 & (\citenum{Druard14}) & 10.16 & (\citenum{Schruba19a}) &  4.78 & (\citenum{Schruba19a}) &  4.48 & (\citenum{Schruba19a}) &  1.97 & (\citenum{Colombo14a}) &  1.89 & (\citenum{Colombo14a}) &  0.89 & (\citenum{Leroy15a}) &  1.19 & (\citenum{Utomo15}) \\
$\Sigma_\mathrm{gmc}$ & \Msunperpc &  650.5 & (\citenum{Henshaw16}) &   98.4 & (\citenum{Rice16}) &   23.4 & (\citenum{Rice16}) &   37.8 & (\citenum{Wong11}) &   27.3 & (\citenum{Druard14}) &   22.4 & (\citenum{Schruba19a}) &   20.1 & (\citenum{Schruba19a}) &   32.1 & (\citenum{Schruba19a}) &  302.2 & (\citenum{Colombo14a}) &  225.9 & (\citenum{Colombo14a}) & 9836.3 & (\citenum{Leroy15a}) & 1479.6 & (\citenum{Utomo15}) \\
$\tau_\mathrm{cross}$ & Myr &  0.22 & (\citenum{Henshaw16}) &  9.35 & (\citenum{Rice16}) & 11.32 & (\citenum{Rice16}) &  6.01 & (\citenum{Wong11}) & 10.55 & (\citenum{Druard14}) &  4.98 & (\citenum{Schruba19a}) &  7.17 & (\citenum{Schruba19a}) &  6.26 & (\citenum{Schruba19a}) &  4.26 & (\citenum{Colombo14a}) &  4.84 & (\citenum{Colombo14a}) &  0.80 & (\citenum{Leroy15a}) &  1.36 & (\citenum{Utomo15}) \\
$\tau_\mathrm{ff}$ & Myr &  0.44 & (\citenum{Henshaw16}) &  7.12 & (\citenum{Rice16}) & 15.76 & (\citenum{Rice16}) &  7.48 & (\citenum{Wong11}) & 15.04 & (\citenum{Druard14}) & 11.59 & (\citenum{Schruba19a}) & 11.27 & (\citenum{Schruba19a}) &  8.79 & (\citenum{Schruba19a}) &  4.00 & (\citenum{Colombo14a}) &  4.83 & (\citenum{Colombo14a}) &  0.58 & (\citenum{Leroy15a}) &  1.12 & (\citenum{Utomo15}) \\
\cutinhead{\footnotesize Pressure Parameters}
log P$_\mathrm{int}$ & K~cm$^{-3}$ &  7.74 &  &  5.28 &  &  4.53 &  &  4.77 &  &  4.55 &  &  4.89 &  &  4.46 &  &  4.73 &  &  6.46 &  &  6.34 &  &  9.00 &  &  8.03 &  \\
log P$_\mathrm{self}$ & K~cm$^{-3}$ &  6.65 &  &  5.12 &  &  3.95 &  &  4.34 &  &  4.14 &  &  3.84 &  &  3.71 &  &  4.16 &  &  6.11 &  &  5.96 &  &  9.14 &  &  7.82 &  \\
log P$_\mathrm{ext,star}$ & K~cm$^{-3}$ &  7.23 &  &  4.70 &  &  4.18 &  &  4.17 &  &  4.23 &  &  3.79 &  &  3.89 &  &  3.69 &  &  5.74 &  &  4.73 &  &  7.41 &  &  5.18 &  \\
log P$_\mathrm{ext,gas}$ & K~cm$^{-3}$ &  7.43 &  &  3.83 &  &  4.05 &  &  4.10 &  &  3.38 &  &  2.83 &  &  3.60 &  &  3.29 &  &  5.32 &  &  4.62 &  &  6.42 &  &  5.77 &  \\
\enddata
\tablerefs{(\citenum{Amblard14}) \citealt{Amblard14} (\citenum{Barnes17}) \citealt{Barnes17} (\citenum{BlandHawthorn16}) \citealt{BlandHawthorn16} (\citenum{Bovy13}) \citealt{Bovy13} (\citenum{Braun09}) \citealt{Braun09} (\citenum{CalduPrimo13}) \citealt{CalduPrimo13} (\citenum{Ciardullo02}) \citealt{Ciardullo02} (\citenum{Dalcanton12}) \citealt{Dalcanton12} (\citenum{Davis14}) \citealt{Davis14} (\citenum{Dorman15}) \citealt{Dorman15} (\citenum{Druard14}) \citealt{Druard14} (\citenum{Ford13}) \citealt{Ford13} (\citenum{Gieren04}) \citealt{Gieren04} (\citenum{Gratier10b}) \citealt{Gratier10b} (\citenum{Henshaw16}) \citealt{Henshaw16} (\citenum{Heyer15}) \citealt{Heyer15} (\citenum{Hughes13b}) \citealt{Hughes13b} (\citenum{Iodice14}) \citealt{Iodice14} (\citenum{Jameson16}) \citealt{Jameson16} (\citenum{Kalberla09}) \citealt{Kalberla09} (\citenum{Kam15}) \citealt{Kam15} (\citenum{Kang16}) \citealt{Kang16} (\citenum{Kennicutt12}) \citealt{Kennicutt12} (\citenum{Kruijssen14b}) \citealt{Kruijssen14b} (\citenum{Launhardt02}) \citealt{Launhardt02} (\citenum{Leroy08}) \citealt{Leroy08} (\citenum{Leroy15a}) \citealt{Leroy15a} (\citenum{Licquia15}) \citealt{Licquia15} (\citenum{Licquia16}) \citealt{Licquia16} (\citenum{Longmore13}) \citealt{Longmore13} (\citenum{Lucero13}) \citealt{Lucero13} (\citenum{Lucero15}) \citealt{Lucero15} (\citenum{Meidt13}) \citealt{Meidt13} (\citenum{MunozMateos07}) \citealt{MunozMateos07} (\citenum{MunozMateos15}) \citealt{MunozMateos15} (\citenum{Nakanishi16}) \citealt{Nakanishi16} (\citenum{Nieten06}) \citealt{Nieten06} (\citenum{Pety13}) \citealt{Pety13} (\citenum{Rekola05}) \citealt{Rekola05} (\citenum{Rice16}) \citealt{Rice16} (\citenum{Sanders03}) \citealt{Sanders03} (\citenum{Schruba10}) \citealt{Schruba10} (\citenum{Schruba12}) \citealt{Schruba12} (\citenum{Schruba19a}) \citealt{Schruba19a} (\citenum{Schruba19b}) \citealt{Schruba19b} (\citenum{Schuster07}) \citealt{Schuster07} (\citenum{Seigar11}) \citealt{Seigar11} (\citenum{Skibba12}) \citealt{Skibba12} (\citenum{Sofue16}) \citealt{Sofue16} (\citenum{StaveleySmith03}) \citealt{StaveleySmith03} (\citenum{Tamm12}) \citealt{Tamm12} (\citenum{Tonry01}) \citealt{Tonry01} (\citenum{Utomo15}) \citealt{Utomo15} (\citenum{Verley09}) \citealt{Verley09} (\citenum{Westmeier11}) \citealt{Westmeier11} (\citenum{Wong09}) \citealt{Wong09} (\citenum{Young11}) \citealt{Young11} (\citenum{vanderMarel02}) \citealt{vanderMarel02} }
\end{deluxetable*}
\setlength{\tabcolsep}{6pt} 
\renewcommand{\arraystretch }{1.2} 

\subsection{Galaxy Properties}
\label{sec:galaxies}

Our galaxy sample comprises eight local galaxies spanning a wide range of stellar mass and morphology. We include one dwarf irregular: the LMC, two low-mass spirals: M33 and NGC300, two massive spirals: the Milky Way and M51, one starburst galaxy center: NGC253, one green-valley galaxy: M31, and one lenticular galaxy: NGC4526. 

We compile global measurements of the stellar mass ($M_\star$), atomic ($M_{\rm atom}$) and molecular gas mass ($M_{\rm mol}$), and star formation rate (SFR) from the literature or we derive them from our own data. Our sample covers about two orders of magnitude in various host galaxy properties: $M_\star = 2 {\times} 10^9 - 10^{11}$ \Msun, ${\rm SFR} = 0.03 - 4.2$ \Msunperyr, ${\rm sSFR} = 2 {\times} 10^{-12} - 2 {\times} 10^{-10}$ yr$^{-1}$, and $f_{\rm gas} = M_{\rm gas} / {(M_{\rm gas} + M_\star)} = 0.01 - 0.5$.

We subdivide our galaxies into local galactic environments, separating central regions, inner disks, and outer disks. These different regions have different surface densities and ISM properties, which may affect the cloud properties and star formation efficiency. Table~\ref{t1} presents the local environments for each target. 

For each environment, we determine the local galactic properties at the median galactocentric radius of the respective cloud sample (see below). We use radial profiles of each tracer, with their total flux pinned to the global properties, to determine the surface densities of stars ($\Sigma_\star$), atomic ($\Sigma_{\rm atom}$) and molecular gas ($\Sigma_{\rm mol}$), star formation rate ($\Sigma_{\rm sfr}$), and the (vertical) velocity dispersions of the stellar and gaseous disks ($\sigma_\star$ and $\sigma_{\rm gas}$).

The range in local galactic properties is even broader than the range of global ones. Our sample covers $\Sigma_\star = 25-4000$ \Msunperpc, $\Sigma_{\rm atom} = 3-18$ \Msunperpc, $\Sigma_{\rm mol} = 2-1000$ \Msunperpc, $\Sigma_{\rm sfr} = 1.5-3500$ \Msunperpc\ Gyr$^{-1}$, and molecular depletion time $\tau_{\rm dep} = 0.2-4.4$ Gyr. From these local galactic properties we derive the ambient gas (midplane) pressure, $P_{\rm ext}$ as defined in Section~\ref{sec:virialtheorem}.

\subsection{Cloud Properties}
\label{sec:clouds}

These extragalactic surveys approach the detail of the CfA-Chile survey of the Milky Way \cite[][${\sim}30$ pc at the Galactic Ring]{Dame01}, typically with sensitivity to clouds with molecular gas mass as low as $\sim 10^4 - 10^5$~M$_\odot$. They also use different telescopes and achieve different physical resolution and sensitivity, and in some cases use different tracers of the molecular gas. As a result, our combined data set is heterogeneous in nature.

Whenever we had access to the original data sets, we extracted the measurements ourselves. We use an updated version of the CPROPS code\footnote{\href{https://github.com/akleroy/cpropstoo}{\tt https://github.com/akleroy/cpropstoo}} \citep{Rosolowsky06, Leroy15a}. The code identifies emission peaks with signal-to-noise $\geq 5$ in two adjacent channels. Then it assigns adjacent pixels to an emission peak until an intensity level is reached where the pixel cannot be uniquely assigned to one peak. Finally, the properties of a cloud are determined from the intensity-weighted moments. These are corrected for the effects of blending and finite sensitivity assuming that the cloud can be approximated as a three dimensional Gaussian. Then the measured size and line width are corrected for the intrinsic spatial and spectral resolution of the data. 

When the original data were not available or re-extraction was not feasible, we adopt the literature measurements scaled to our assumed distances.

We define the macroscopic cloud properties as follows:
\begin{align}
&{\it Radius:} &\quad R &= 1.91 \sqrt{\sigma_x \sigma_y} \label{eq:rad}\\[1.5ex]
&{\it Luminous~Mass:} &\quad M &= \alpha_{\rm CO} L_{\rm CO} \\
&{\it Virial~Mass:} &\quad M_{\rm vir} &= \dfrac{5 \sigma_{\rm v}^2 R}{G} \\
&{\it Surface~Density:} &\quad \Sigma &= 0.77 \dfrac{M}{\pi R^2} \\
&{\it Volume~Density:} &\quad \rho &= 1.26 \dfrac{M}{4/3 \pi R^3} \label{eq:vol}
\end{align}
\noindent We note that all masses and densities in this paper include the contribution of heavy elements. $\sigma_x$ and $\sigma_y$ in Eq.~\eqref{eq:rad} denote the size of a cloud (defined by the intensity-weighted second moment along the two spatial axes).

From these macroscopic cloud properties we derive the cloud's free-fall time, $\tau_{\rm ff}$, and turbulent crossing time, $\tau_{\rm cross} = R / {\sqrt{3} \sigma_{\rm v}}$ (for one-dimensional velocity dispersions), as well as the cloud's internal turbulent pressure, $P_{\rm int}$, and the self-gravitational pressure, $P_{\rm self}$, as defined in Eqs.\ \mbox{\ref{eq:Pint}$-$\ref{eq:Pself}}.

We note that the definitions of the macroscopic cloud properties (Eqs.~\ref{eq:rad}$-$\ref{eq:vol}) and the definitions of the cloud virial parameter and internal pressures (Eqs.~\ref{eq:avirobs}$-$\ref{eq:Pself}) bear minor inconsistencies in the geometric factors describing the density distribution of the molecular clouds, i.e., constant, $1/r$, or Gaussian density profiles. For now we accept this to match common definitions and account for these differences as part of the systematic uncertainties (see below). In the future we intend to remedy this inconsistency by determining the geometric factors from higher dynamic range data obtained with ALMA.

To derive molecular cloud masses, we have adopted the standard Galactic CO\mbox{(1-0)}-to-H$_2$ conversion factor: $\aco = 4.35$ \acounits\ (which includes a correction for heavy elements) for the Milky Way disk, M31, M51, NGC4526; for the low mass galaxies LMC, M33, and NGC300 we adopt twice this value \citep[][]{Wong11, Druard14, Kruijssen19b}. For M33 we further adopt a CO\mbox{(2-1)}-to-CO\mbox{(1-0)} brightness temperature ratio of $0.8$ \citep{Druard14}. For the Milky Way center and NGC253 molecular gas masses are derived from optically-thin dense gas tracers and dust continuum \citep[][]{Leroy15a, Walker15}.

The above method to calculate cloud properties remedies as best as possible the signatures of inhomogeneous resolution and sensitivity between our different data sets. Despite this effort, we note that peak identification methods such as CPROPS naturally tend to identify structures with size similar to the native resolution of the data set \citep[Table~\ref{t1}; see also][]{Leroy16}.

For each of the local galactic environments defined in Section~\ref{sec:galaxies} and Table~\ref{t1}, we derive mass-weighted average cloud properties. We note that for the calculation of the mass-weighted averages the order in which cloud properties and averages are calculated matters. We calculate all macroscopic cloud properties (Eqs.\ \mbox{\ref{eq:rad}${-}$\ref{eq:vol}}), the pressure related quantities and thereof derived `observed' and `predicted' virial parameter (Eqs.\ \mbox{\ref{eq:avirobs}${-}$\ref{eq:Pext}}) for each cloud individually and then determine their mass-weighted average. We have tested whether our results depend on this methodology by calculating the virial parameter and pressure quantities from mass-weighted average cloud properties instead and find no major difference.

This sample-averaging resembles the methodology described in \citet{Leroy16} and distinguishes our work from previous studies. Appropriate averaging over all clouds in a galactic environment highlights the impact of environment on the mean cloud properties. Previous studies have plotted the entire cloud population, often emphasizing offsets among scaling relations. This shows the full dynamic range of cloud properties \citep[that is frequently interpreted to reflect the time evolution of the clouds; e.g.,][]{Murray11, LeeE16, Padoan17} but can obscure the dependence of cloud properties and star formation on environment.

Our study also expands the range of galactic environments for which cloud properties and star formation models have been tested. Most work so far has focused on single targets, either the Milky Way \citep{Field11, Murray11, Vutisalchavakul16, Barnes17} or individual selected local galaxies \citep{Utomo15, Leroy17b, Ochsendorf17, Schruba17}. The  best synthetic work today, by \citet[][]{Hughes13b}, had only three high quality data sets available. 

\subsection{Uncertainties and Error Propagation}
\label{sec:uncertainties}

We account for uncertainties in the cloud properties, their sample averages, and the galactic disk parameters. We consider (a) statistical uncertainties due to finite signal-to-noise (S/N) and uncertain sensitivity corrections, (b) statistical uncertainties due to sample variance when determining the mass-weighted averages, and (c) systematic uncertainties in the mass-to-light conversion factors and the adopted geometric factors.

Because the statistical uncertainties in the cloud properties are not always available (from the literature), we adopt typical uncertainties derived from in-hand data sets. We estimate these based on our application of the CPROPS code to cases where we have the original data cube \citep[including extrapolations to the zero noise level; see][and the CPROPS documentation]{Rosolowsky06, Leroy15a}. For our typical case of marginally resolved clouds detected at peak ${\rm S/N} \sim 5{-}10$, we find logarithmic uncertainties of $0.1$~dex for $R$; $0.15$~dex for $\sigma_{\rm v}$; $0.25$~dex for $M_{\rm vir}$; $0.3$~dex for $M_{\rm lum}$, $\Sigma$, $\tau_{\rm ff}$; and $0.35$~dex for $\alpha_{\rm vir}$.

We derive uncertainties in the (mass-weighted) averages by a Monte-Carlo analysis which accounts for (i) sample variance by bootstrapping, and (ii) statistical uncertainties by perturbing the cloud properties. In this work, we do not study resolution or completeness effects but work with the best current data at their native resolution. We plan to investigate these effects in the future. See \citet{Sun18} for more analysis of the possible impact of these factors.

Many of our calculations rely on knowledge of the mass of gas or stars. These are affected by systematic uncertainty in the light-to-mass conversion. For the CO-to-H$_2$ conversion factor, we adopt an uncertainty of $0.2$~dex for massive galaxies. For smaller galaxies, with metallicity $Z \lesssim 0.5 Z_\odot$ (i.e., LMC, M33, NGC300), we treat the CO-to-H$_2$ conversion factor as uncertain by $0.3$~dex. For the $3.6\,\micron$-to-$M_\star$ factor, we adopt a fixed uncertainty of $0.15$~dex. 

Different assumptions in cloud geometries (e.g., constant, $1/r$, or Gaussian density profiles) lead to additional systematic uncertainties of $0.15$~dex in radius and $0.10$~dex in line width. We analytically (or numerically for the sample averages) propagate the uncertainties from the input parameters to the derived parameters. 

Note that we do not account for the covariance among the uncertainties. This implies that our quoted uncertainties represent conservative upper limits.

In general, we find that the statistical uncertainties on the properties of individual molecular clouds are significantly suppressed by our approach of determining sample averages and often reduce to $\lesssim 0.05{-}0.10$~dex, with the exception of galactic environments with very small cloud catalogs (i.e., NGC253). Therefore, the systematic uncertainties dominate our formal error budget. However, we note that these systematic uncertainties are expected to bias parameters in the same way for many/all galaxies in our sample, and thus should again be interpreted as conservative upper limits on the uncertainties. In the following figures, we show these two types of uncertainties: colored error bars account for all sources of uncertainties except for systematic ones, whereas grey error bars also include systematic uncertainties.

\section{Results}
\label{sec:results}

In Table~\ref{t1} we present our compilation of cloud properties in context for the Milky Way and seven local galaxies. The table includes \emph{global galaxy properties} that describe our galaxy sample and can be used to reference to other samples. The \emph{local galaxy properties} give the galactic disk properties at the median galactocentric radius of our cloud sample; these quantities set the ambient medium pressure and the molecular gas depletion time. The \emph{local cloud properties} list the mass-weighted mean properties of the cloud population and the fraction of molecular mass in each cloud catalog. The \emph{pressure parameters} state the mass-weighted clouds' turbulent and self-gravitational pressures, and the ambient medium pressures by the stellar and gaseous disks. We use these measurements to study the dependence of the clouds' dynamical state on local galactic environment, to assess their star formation efficiency, and to test theories of star formation in turbulent clouds.

\subsection{Dynamical State of Molecular Clouds and its~Dependence on Galactic Environment}
\label{sec:dynstate}

\begin{figure}[tb]
\includegraphics[width=0.48\textwidth]{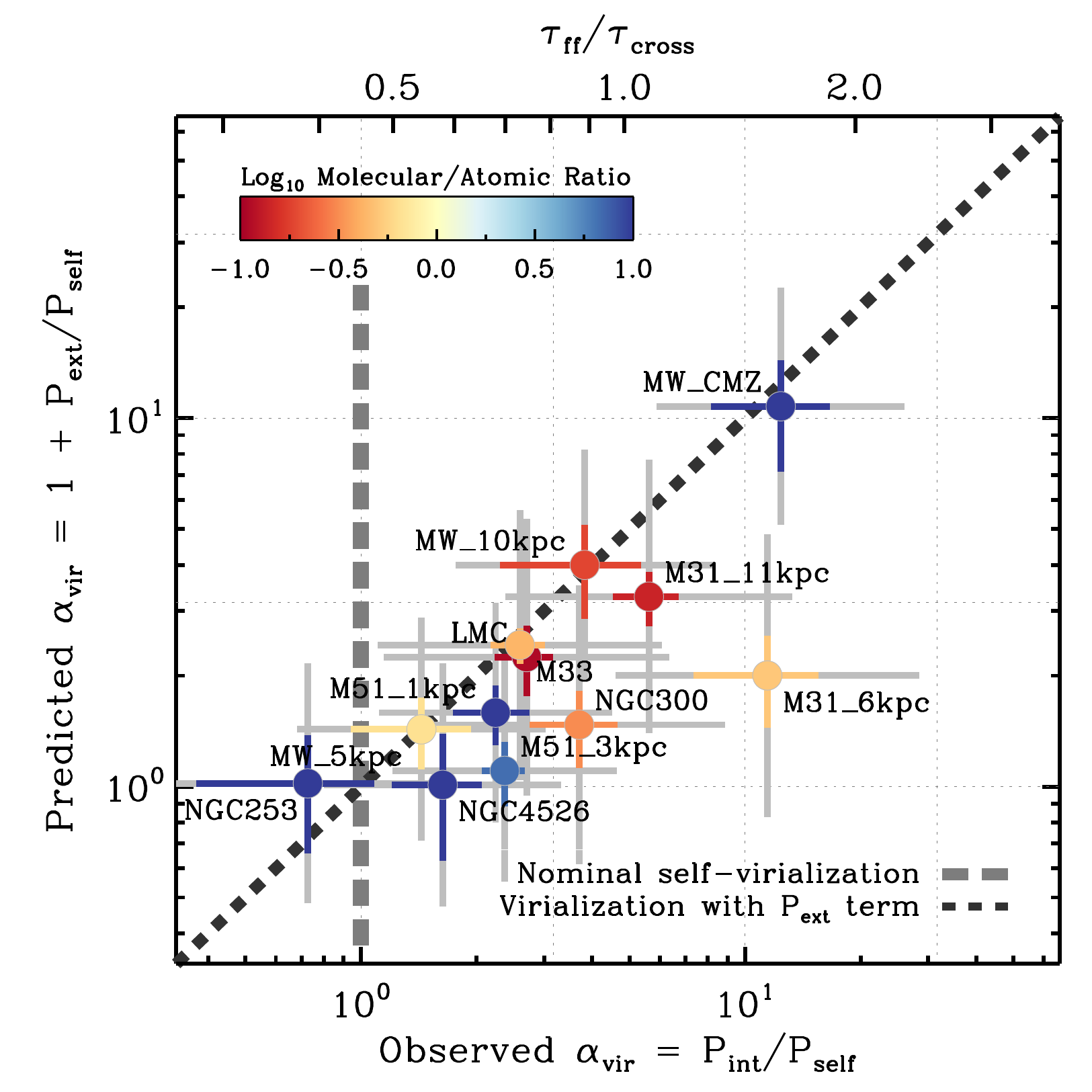}
\caption{Virial parameter, $\alpha_{\rm vir}$, of molecular clouds derived from the observed balance of the clouds' kinetic and self-gravitational pressures ($x$-axis) and as predicted for virialized clouds in external pressure equilibrium ($y$-axis). Data points show the mass-weighted average of the cloud population of an entire galaxy or a distinct subregion therein. Error bars represent the statistical uncertainties and sample variance (in color) and also including systematic uncertainties (in grey). The color-coding shows whether atomic or molecular gas dominates the ISM. Virialized clouds confined only by self-gravity lay along the vertical, dashed line; those confined by external pressure along the diagonal, dotted line.
\label{f1}}
\end{figure}

Figure~\ref{f1} illustrates the relationship between the dynamical state of molecular clouds and local galactic environment. The $x$-axis shows the virial parameter, $\alpha_{\rm vir}$, (Eq.~\ref{eq:avirobs}) derived from the observed balance of the clouds' kinetic and self-gravitational pressures (Eqs.~\ref{eq:Pint}$-$\ref{eq:Pself}). The $y$-axis shows the value of $\alpha_{\rm vir}$ predicted for virialized clouds in external pressure equilibrium (Eq.~\ref{eq:avirtheo}). Data points show the mass-weighted average of the cloud population for each region that we study (Section~\ref{sec:clouds}). Error bars represent the statistical uncertainties and sample variance (in color) and also including systematic uncertainties (in grey;  Section~\ref{sec:uncertainties}).

A main finding of this paper is the close relationship between $\alpha_{\rm vir,\,obs}$ and $\alpha_{\rm vir,\,theo}$. We find a median ratio of ${\sim}0.83$ and scatter of ${\sim}0.3$~dex. In our data, the scaling between $\alpha_{\rm vir,\,obs}$ and $\alpha_{\rm vir,\,theo}$ is stronger than the relationship between internal kinetic pressure, $P_{\rm int}$, and the pressure associated with self-gravity, $P_{\rm self}$. It is also stronger than the scaling between internal kinetic pressure, $P_{\rm int}$, and external pressure $P_{\rm ext}$. In our data, both of those pressure scalings exhibit a dispersion of ${\sim}0.5$~dex. That scatter in the pressure--pressure scalings appears to originate from a systematic trend between $P_{\rm self} / P_{\rm ext}$ with $P_{\rm int}$. 

Put another way, Figure~\ref{f1} implies that molecular clouds in local galaxies are in virial equilibrium \textit{once the confining external pressure is accounted for} (i.e., diagonal dotted line). But they are not necessarily in a simple virial equilibrium set only by the clouds' self-gravity (i.e., vertical dashed line). The local galactic environment has a clear and substantial imprint on the dynamical state of molecular clouds. We find the (sample averaged) virial parameter to have large (factor ${\sim}15$) systematic variation among our galaxies and distinct subregions therein, and we find that these variations are significantly larger than all sources of uncertainty. In some environments, clouds have virial parameters near unity (i.e., the bottom left part of Figure~\ref{f1}). These clouds appear virialized considering only their self-gravity and external pressure has no major impact. In other environments, clouds have virial parameters of ${\sim}3{-}10$ (i.e., the top right part of Figure~\ref{f1}), indicating these clouds have kinetic energies much larger than their self-gravitational energies. These clouds are either unbound and transient, or they are pressure-confined.

\begin{deluxetable}{p{2.2cm}p{2.0cm}p{2.0cm}}[tb]
\tablewidth{0.8\columnwidth} 
\tablecaption{Two Classes of Clouds \& Environments\label{t2}}
\tablecolumns{3}
\tablehead{\colhead{Property} & \colhead{Pressurized} & \colhead{Self-gravitating}}
\startdata
$P_{\rm ext}$ [K~cm$^{-3}$] & $5{\times}10^3 - 3{\times}10^4$ & $4{\times}10^4 - 7{\times}10^7$\\
$\Sigma_{\rm star}$ [\Msunperpc] & $25-170$ & $200-4000$\\
$\Sigma_{\rm gas}$ [\Msunperpc]	& $10-23$ & $17-10^3$\\
$R_{\rm mol}$ & $0.1-0.7$ & $5-200$\\[1ex]
\tableline
$\Sigma_{\rm gmc}$ [\Msunperpc]	& $10-30$ & $10^2 - 10^4$\\
$\alpha_{\rm vir,\,obs}$ & $3-10$ & $0.7-3$\\
\enddata
\tablecomments{The table lists the full range of the environmental (top) and cloud (bottom) properties for the two classes of clouds and galactic environments (see text).}
\end{deluxetable}

These systematic variations in the dynamical state of molecular clouds link to properties of the local galactic environment. We qualitatively break our sample into two classes of clouds, living in two different types of environments. We list the approximate properties of these two classes in Table~\ref{t2}. 

To first order, ``self-gravitating'' clouds have high surface densities ($\Sigma_{\rm gmc} \approx 10^2 {-} 10^4$ \Msunperpc) and are found in high ambient pressure environments ($P_{\rm ext} \approx 4{\times}10^4 - 7{\times}10^7$ K~cm$^{-3}$) as in the central and inner Milky Way, M51, NGC253, and NGC4526. Despite the high ambient pressure in their environments, these clouds appear nearly virialized considering only their self-gravity and internal motions.\footnote{For the Galactic Centre, we only consider clouds on the `dust ridge' of the `100-pc stream' \citep{Walker15}, which contains the highest-density clouds in the region. For clouds at larger galactocentric radii, we expect a larger influence of the external pressure \citep[e.g.,][]{Kruijssen14b}.}

High virial parameter, (externally) pressurized clouds have low surface densities ($\Sigma_{\rm gmc} \approx 10 {-} 30$~\Msunperpc) and reside in low ambient pressure environments ($P_{\rm ext} \approx 5{\times}10^3 - 3{\times}10^4$ K~cm$^{-3}$) as in the outer Milky Way, LMC, M31, M33, and NGC300. These clouds appear virialized only once external pressure terms are taken into account.

The dependence of the clouds' virial parameters on their surface densities is also apparent when considering clouds of different masses. To investigate this, we have considered sub-samples of clouds with masses within logarithmic bins between $10^4$~\Msun\ and $10^8$~\Msun. We find that lower mass clouds ($M_{\rm gmc} = 10^4 {-} 10^6$~\Msun) have high virial parameters ($\avir \approx 2{-}10$), while massive clouds ($M_{\rm gmc} = 10^6 {-} 10^8$~\Msun) have low virial parameters ($\avir \approx 1{-}3$). One can thus say that the more massive a cloud is, the more likely it is to have high surface density and therefore decouple dynamically from its environment. However, the threshold when this decoupling sets in is a strong function of the density of clouds' ambient medium \citep[see also][and in prep.]{Meidt18}.

The color coding in Figure~\ref{f1} highlights one major difference distinguishing the environments where the two classes of clouds are found. Color indicates whether the ISM is predominantly composed of atomic or molecular gas. The figure shows that the ``pressurized'' clouds are more likely to be found in parts of galaxies where atomic gas makes up most of the neutral interstellar medium.

Such systematic variations in cloud properties with environment have been suggested by molecular cloud surveys in the inner and outer Milky Way \citep{Heyer01, Heyer09, Field11} and seen contrasting M33, M51, and the LMC by \citet{Hughes13a, Hughes16}. Here, we roughly triple the sample of galaxies studied and synthesize Milky Way measurements and extragalactic work. This larger data set clearly shows that the dynamical state of molecular clouds depends on environment in a systematic way.

Our results also agree with recent work by \citet{Leroy16} and \citet{Sun18}. Those papers present a similarly large dynamic range in the virial parameter of molecular gas in nearby galaxies. They adopt the pixel-based analysis method developed by \citet{Leroy16} and measure the gas surface density and line width at fixed spatial scales of ${\sim}45{-}120$~pc, which with an assumption on the beam filling factor and the gas extent along the line of sight treces the virial parameter. They also find that molecular gas in massive, molecular gas dominated spiral galaxies has $\alpha_{\rm vir} \sim 2$ (i.e., energy equipartition) while the gas-poor, atomic gas dominated galaxies M31 and M33 have $\alpha_{\rm vir} \sim 3{-}10$. This agreement is partially by construction, because \citet{Leroy16} and \citet{Sun18} analyze the same M31, M33, and M51 data that we consider here.

\citet{Sun18} suggest several possible drivers for the high apparent $\alpha_{\rm vir}$ in M31 and M33 including (a) beam dilution of molecular clouds much smaller than their $45{-}80$~pc measurement scale or (b) the impact of the ambient pressure from the interstellar medium (as we argue here). In this work, we require emission peaks to be (at least marginally) spatially resolved to directly measure their size, line width, and surface densities. Our CPROPS methodology has its own biases, as it tends to find beam sized objects and leads to incompleteness in the measurements. But our measurements offer a strong indication that even resolved clouds in these galaxies appear ``pressurized,'' so that beam dilution is not the sole reason for the high $\alpha_{\rm vir}$ values. We argue that these high virial parameters are indeed a sign of pressure confinement by the diffuse ambient medium in atomic-gas dominated galaxies like M31, M33, NGC300, and the LMC.

Moreover, Figure~\ref{f1} provides substantial evidence in favor of the hypothesis by \citet{Elmegreen89} that the cloud's dynamical state is linked to the ambient gas pressure set by vertical hydrostatic equilibrium (Eq.~\ref{eq:Pext}). However, our definition of $P_{\rm ext}$ differs from \citeauthor{Elmegreen89}'s one by a constant\footnote{We do not reduce the total midplane gas pressure by the relative contributions from cosmic rays and magnetic fields, for which \citet{Elmegreen89} assumed relative contributions of $0.40$ and $0.25$, respectively. Instead, we assume the vertical scale height of cosmic rays and magnetic fields to be larger than the scale height of the neutral gas, such that they have minimal stabilizing effect on the weight of the neutral gas disk.} and in only considering the diffuse ISM to provide the ambient gas pressure in the midplane {(Eq.~\ref{eq:sdism})}\footnote{The diffuse ISM mass is defined as the total neutral gas mass minus the gas mass within molecular clouds, i.e., we assume a minor contribution of molecular clouds to the galactic disk potential \citep[e.g.,][]{Ostriker10}.}. A number of alternative expressions for the ambient pressure confining molecular clouds have been proposed \citep[][and others]{Chieze87, Bertoldi92, Wolfire03, Blitz06, Koyama09b, Ostriker10, Kim11}. These expressions differ in assumptions on the structure of the ISM, the size of atomic shielding layers around molecular clouds, the effect of magnetic fields, and the scale height of cosmic rays; leading to predictions of the ambient gas pressure that differ by a factor of a few. Unfortunately, these differences are comparable to the systematic uncertainties in our measurements of the cloud and environmental parameters, which precludes a firm conclusion as to which expression matches our observations best \citep[cf.][]{Hughes16}. We will investigate this topic in more detail using sensitive cloud-scale mapping of the molecular gas obtained by the PHANGS-ALMA survey.

\subsection{Star Formation Efficiency of Molecular Clouds\\ and Whole Galaxies}
\label{sec:sfedyn}

We also infer the star formation rate and molecular gas depletion time for each region studied. By comparing this to the average cloud properties, we can compare these against measurements of the star formation efficiency (SFE) on galactic scales and predictions of star formation theories in the literature (Section~\ref{sec:comparison}).

\begin{figure}[tb]
\includegraphics[width=0.48\textwidth]{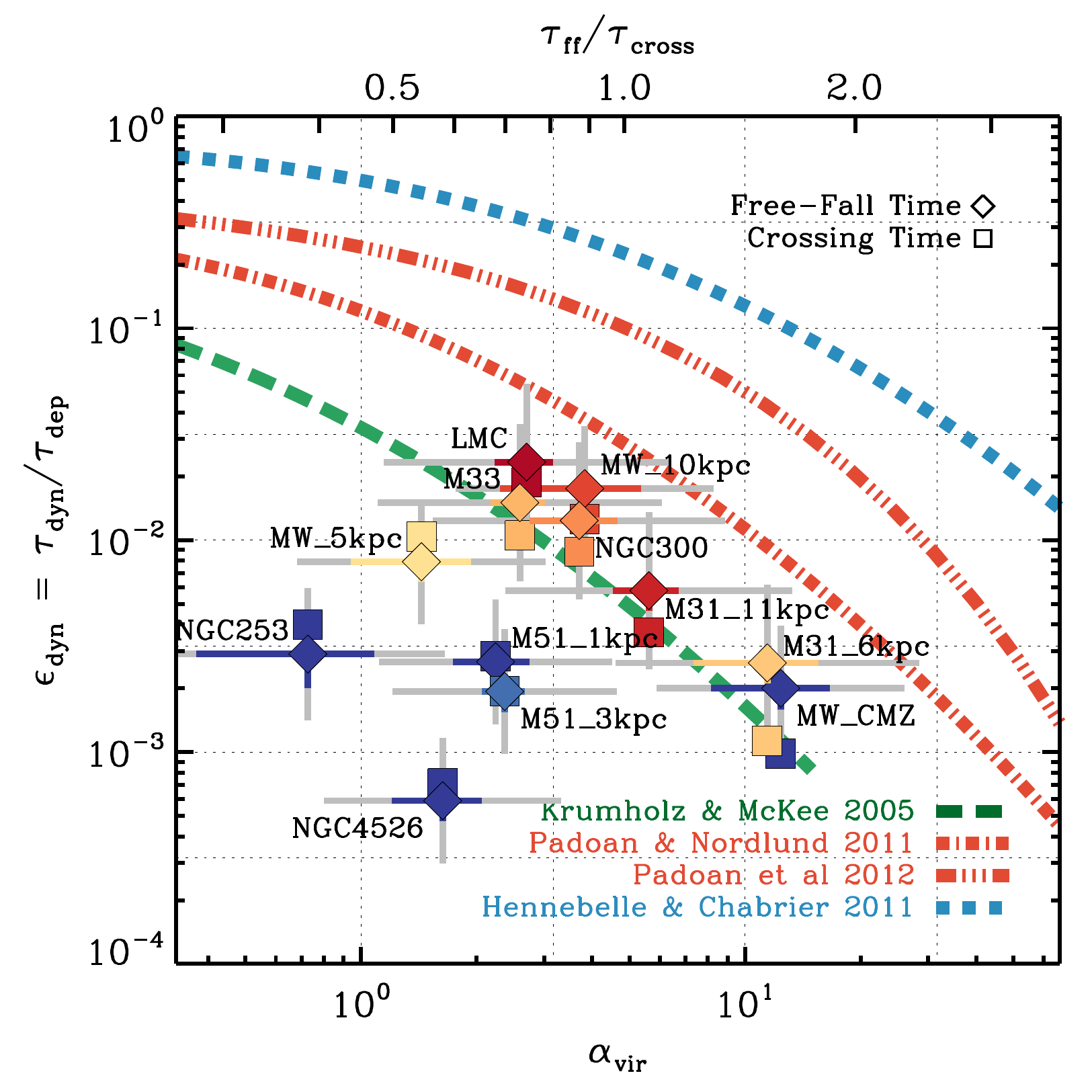}
\caption{Star formation efficiency, $\epsilon_{\rm dyn}$, as function of the virial parameter, $\alpha_{\rm vir}$, and the related ratio of free-fall and turbulent crossing time, $\tau_{\rm ff} / \tau_{\rm cross}$ (top x-axis). Data points show the mass-weighted average of the cloud population of an entire galaxy or a distinct subregion therein. Error bars represent the statistical uncertainties and sample variance  (in  color)  and  also  including  systematic  uncertainties (in grey).  The color-coding shows whether atomic or molecular gas dominates the ISM (as in Figure~\ref{f1}). Dashes lines show theoretical predictions for the scaling of $\epsilon_{\rm dyn}$ with $\alpha_{\rm vir}$ or $\tau_{\rm ff} / \tau_{\rm cross}$, respectively (see text).
\label{f2}}
\end{figure}

Figure~\ref{f2} shows the SFE per dynamical time, $\epsilon_{\rm dyn}$, as a function of the virial parameter, $\alpha_{\rm vir}$. 
$\epsilon_{\rm dyn} = \tau_{\rm dyn} / \tau_{\rm dep}$ is the ratio of the local dynamical timescale, $\tau_{\rm dyn}$, to the global molecular gas depletion timescale, $\tdep = M_{\rm mol} / {\rm SFR}$ determined over the range of galactocentric radii of the cloud populations. Thus, $\epsilon_{\rm dyn}$ expresses the fraction of molecular gas converted to stars per dynamical timescale. $\epsilon_{\rm dyn}$ represents a slight generalization of the \eff\ discussed in Section~\ref{sec:theories}, which treated the gravitational free-fall time as $t_{\rm dyn}$.

In Figure~\ref{f2}, each data point shows the mass-weighted average of the cloud population for one of our regions. For each region, we show two data points that consider different local dynamical timescales, $\tau_{\rm dyn}$, i.e., either the free-fall time, $\tau_{\rm ff}$ (diamond symbols), or the turbulent crossing time, $\tau_{\rm cross}$ (square symbols). For the points adopting $\tau_{\rm dyn} = \tau_{\rm ff}$, we also show the statistical uncertainty including sample variance (colored error bars) and the full uncertainty range including also systematic uncertainty (grey error bars). The uncertainties for the data points including \tcross\ are comparable, but omitted for clarity. As in Figure~\ref{f1}, data points are color-coded to indicate whether atomic or molecular gas dominates the ISM. All data values shown here can be found in Table~\ref{t1}. The top axis of the figure re-expresses the virial parameter, \avir, as ${(\tff / \tcross)}^2 = \pi^2 \Gamma \avir/8$, adopting our fiducial $\Gamma = 0.6$.

\emph{Low SFE per dynamical time:} Figure~\ref{f2} shows that (i) \edyn\ varies between $0.05\% - 3\%$, (ii) these variations are much larger than any source of uncertainty (and are therefore real), and (iii) there is no strong difference whether \tdyn\ is set to \tff\ or \tcross. The values of \edyn\ are low in an absolute sense, with only $0.1\%-1\%$ of the molecular gas mass converted to stars per collapse time. They are in good agreement with previous studies connecting cloud-scale \tff\ to the disk-averaged \tdep\ in the Milky Way \citep{Murray11, Vutisalchavakul16}, in M51 \citep{Leroy17b}, or in a sample of massive spiral galaxies \citep{Utomo18}. Our measurements are also in good agreement with studies of the kpc-scale distribution of dense gas, bulk molecular gas, and recent star formation in samples of nearby galaxies \citep{GarciaBurillo12, Usero15}. They are often but not always lower than the typical $\eff \sim 0.5\%-2\%$ found by studies focused on Galactic star-forming clouds \citep{Evans14, Heyer16, LeeE16, Barnes17}---a discrepancy that we attribute to their selection bias on currently star-forming clouds while neglecting quiescent clouds that dominate the molecular gas budget on Galactic scales \citep[see section \mbox{3.2.3} in][]{Leroy17b}. Our results are significantly (${\sim}25{\times}$) smaller than the recent measurement of $\eff \sim 5\% - 25\%$ for star-forming clouds in the LMC \citep{Ochsendorf17}. The underlying cause for this difference remains unclear, however, the short global molecular gas depletion time of the LMC \citep[$\tdep \approx 0.4$~Gyr;][]{Jameson16} likely plays a role.

\emph{Scatter and systematic trend in the SFE per dynamical time:} In addition to overall low $\eff$, we also find nearly two orders of magnitude large dynamic range in \eff. Despite this large scatter in our full sample, our previous differentiation between clouds in low and high-pressure environments (or equivalently between atomic or molecular gas dominated regions as highlighted by the color coding of data points) suggests that molecular clouds in low-pressure, atomic-dominated regions (the orange-red symbols) follow a common trend between \edyn\ and \avir\ such that regions with with small virial parameter ($\avir \sim 1{-}3$) have the largest observed SFE per free-fall time ($\eff \sim 1\% - 3\%$), while clouds with large virial parameter ($\avir \sim 5{-}10$) have systematically lower SFE per free-fall time ($\eff \sim 0.1\% - 0.7\%$). For the other class of clouds living in high-pressure, molecular-dominated regions (the blue symbols), we observed among the lowest values for the SFE per free-fall time ($\eff \sim 0.05\% - 0.5\%$) and find no clear trend with the clouds' virial parameters.

A large dynamic range in \eff\ has been measured studying individual clouds in the Milky Way and the LMC \citep{LeeE15, Murray15, Ochsendorf17}. These variations have been interpreted as the evolution of individual clouds. But our measurements average over large areas and many molecular clouds in distinct evolutionary states. They must reflect systematic differences among our targets.

\emph{Free-fall or crossing time as fiducial timescale of gravitational collapse and star formation:} The models of turbulence regulated star formation discussed in Section~\ref{sec:theories} motivate the free-fall time to be the relevant time scale for star formation. While we do find \tdep\ to be correlated with \tff, we note that the correlation of \tdep\ with \tcross\ is equally strong. This finding paired with the large scatter in \eff, does not support the view that the gravitational free-fall time \emph{alone} is a reliable predictor of the star formation rate of real molecular clouds in the disk of galaxies---in addition \tcross, \avir, and other physical parameters see to play an important role. The common assumption (especially in numerical simulations) of a constant SFR per free-fall time of order $\eff \sim 1\%-10\%$ with small $0.4$~dex scatter \citep[e.g.,][]{Krumholz07} does not seem to hold universally across all galactic environments present in the nearby galaxy population. As discussed above, this conclusion agrees with work studying individual systems, which have raised similar concerns about the predictive power of \tff .

\emph{Role of Diffuse Molecular Gas:} Our calculation of \edyn\ depends on the ability to compare the integrated molecular gas depletion time to timescales derived from the resolved cloud population. This could potentially be complicated by any non-star-forming, diffuse molecular gas. Evidence for diffuse molecular gas comes from studies of the spatial and spectral distribution of CO emission in galaxies which estimate it to account for $ \lesssim 50\%$ of the total molecular gas \citep[e.g.,][see also \citealt{Hygate19} proposing a new, physically motivated method to identify diffuse emission that is unrelated to the current star formation process]{Pety13, CalduPrimo13, CalduPrimo16, RomanDuval16}. 

We derive \tdep\ from the total molecular gas mass, $M_{\rm mol}$, as is commonly done. If the molecular gas not in our cloud catalog is indeed in a diffuse phase that is not (immediately) participating in the star formation process, then the instantaneous depletion time for the clouds that we study will be shorter. $M_{\rm mol}$, and so \tdep , would be scaled down on average by a factor $f_{\rm gmc} \sim 0.3$. \edyn\ would be scaled upwards by a factor of ${\sim}3$. Applying such a correction would approximately conserve the difference between low- and high-\avir\ targets, since the relative shift between the two is a factor of $\lesssim 1.5$, so that our main conclusions remain unchanged. Such a correction might help reconcile our observations with some of the Milky Way work, but without understanding the interaction of GMCs and diffuse gas, the physical meaning of such a correction is unclear (e.g., do the two represent different parts of a cloud's life cycle or distinct, long-lived phases).

\subsection{Comparison to Theoretical Models}
\label{sec:comparison}

Figure~\ref{f2} shows the ``single free-fall time'' version of the \citetalias{Krumholz05}, \citetalias{Padoan11}, \citetalias{Padoan12}, and \citetalias{Hennebelle11} models (introduced in Section~\ref{sec:theories}) as dash-dotted curves. Overall, the models have the tendency to predict \edyn\ larger than what we observe, sometimes overpredicting the observations by ${\sim}1{-}2$ orders of magnitude. The discrepancy is largest for clouds in high-pressure environments with small $\avir \approx 1{-}2$ (i.e., the inner, gas-rich disks of galaxies) that have $\edyn \approx 0.05\% {-} 0.5\%$ which is ${\sim}2$~dex below the model predictions. For clouds in low-pressure environments---for which we already noted the finding of an anti-correlation between \edyn\ and \avir---the \citetalias{Krumholz05} (see also footnote~\ref{footnote3}) and \citetalias{Padoan12} models are in good agreement with our observations, whereas the \citetalias{Padoan11} and \citetalias{Hennebelle11} models predict a similar trend but with \edyn\ values that are ${\sim}1$~dex higher than observed. What is apparent from Figure~\ref{f2} is that none of the models can match all our measurements, i.e., clouds in low and high-pressure environments, by a single relationship.

\begin{figure*}
\begin{center}
\includegraphics[width=0.8\textwidth]{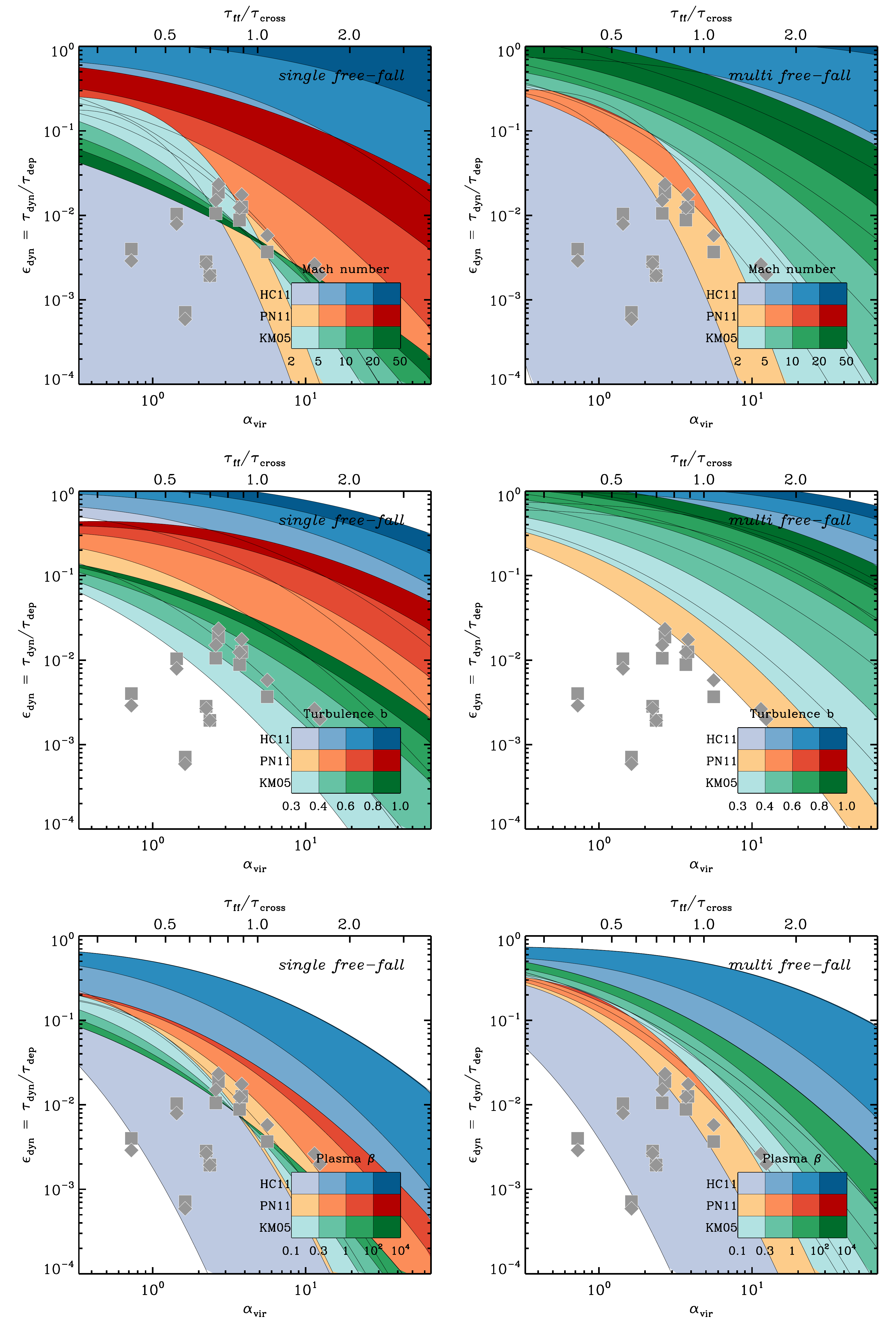}
\end{center}
\caption{Star formation efficiency, \edyn, as function of the observed virial parameter, \avir, and the related ratio of free-fall and turbulent crossing time, $\tau_{\rm ff} / \tau_{\rm cross}$ (top x-axis) (same data as in Figure~\ref{f2}) for the single-free-fall (left panels) and multi-free-fall (right panels) models. The three panels from top to bottom show how the models vary within (plausible) ranges of the unconstrained model parameters: sonic Mach number, $\mathcal{M}$, turbulence driving parameter, $b$, and the plasma $\beta$ parameter describing the strength of the turbulent magnetic field ($\beta \rightarrow \infty$ implies no magnetic fields). In each panel, we vary one of the three parameters while keeping the other two fixed at their default values (see text).
\label{f3}}
\end{figure*}

Figure~\ref{f3} investigates whether the additional model parameters ($\mathcal{M}$, $b$, and $\beta$) can be adjusted to improve the agreement with our observations (as shown in  Figure~\ref{f2}). In each panel, we show the SFE per dynamical time, \edyn, predicted by the \citetalias{Krumholz05}, \citetalias{Padoan11}, and \citetalias{Hennebelle11} models\footnote{The \citetalias{Padoan12} model is a simplification of \citetalias{Padoan11} and does not include dependencies on additional model parameters and is therefore omitted here.} as function of the virial parameter, \avir . In each panel, we vary one of the three additional model parameters. The left column shows the model predictions for the single free-fall time formulation and the right column for the multi free-fall time ones. From top to bottom, the varied model parameters are: the turbulent Mach number ($\mathcal{M}$), the turbulence driving parameter $b$, and the plasma $\beta$ describing the strength of the turbulent magnetic field ($\beta \rightarrow \infty$ for negligible magnetic fields). While varying one of these parameters, we keep the other two fixed at their default values; the considered ranges are $\mathcal{M} = 2{-}50$, $b = 0.3{-}1.0$, $\beta = 0.1{-}10^4$, which represent the maximum plausible ranges for the here studied galactic environments \citep[e.g.,][]{Federrath12,Padoan14}.

We find that the turbulence regulated models of star formation (with the exception of the \citetalias{Hennebelle11} model) struggle to match the observed systematic variations in SFE per dynamical time for any plausible range in their four model parameters (\avir, $\mathcal{M}$, $b$, or $\beta$) without an \emph{ad hoc} adjustment of the overall normalization of the models. We can rule out that variations in \avir\ or $\beta$ alone can reproduce the spread in the observed \edyn\ values. Variations in $\mathcal{M}$ and $b$ cover a (somewhat) larger range of \edyn\ but still smaller than the observed range in \edyn. Thus, simultaneous variations in several parameters would be needed for the models to match the observations. This would require gas-rich, inner galaxy disks (commonly having $\avir \approx 1-2$) to have low Mach number ($\mathcal{M} \sim 2{-}5$), predominantly solenoidal turbulence driving ($b \sim 0.3{-}0.4$), or non-negligible turbulent magnetic pressure ($\beta < 10$). On the other hand, gas-poor, outer galaxy disks (having $\avir \approx 3-20$) would require high Mach number ($\mathcal{M} \sim 10{-}50$), mostly compressive turbulence driving ($b \sim 0.5{-}0.8$), and negligible turbulent magnetic pressure ($\beta > 100$). These parameter ranges represent predictions for future observations, under the assumption that current turbulent star formation theories accurately describe the star formation rates in the environments considered.

Each of the above conditions must be satisfied in order to reconcile the turbulent cloud models with our observations. Encouragingly, it is reasonable to expect an increased degree of solenoidal turbulence driving \citep[due to shear, e.g.][]{Krumholz15b, Kruijssen19a} or elevated magnetic pressures \citep[e.g.,][]{Pillai15, Federrath16} towards galactic centres. However, the requirement of a low Mach number in the inner disks of galaxies stands against flat or falling radial velocity dispersion profiles observed in nearby galaxies \citep[e.g.,][]{Wilson11, CalduPrimo13, Mogotsi16, Sun18} or simple disk-center decompositions carried out for the Milky Way \citep[e.g.,][]{Oka01,Shetty12,Kruijssen13}. This means that the data points at low $\avir$ and low $\eff$ (bottom left in Figure~\ref{f3}) are out of reach of the considered models.

Overall, we conclude that if one searched for a description of the SFE per free-fall time that works in all environments found in nearby galaxies, then the simplest model of a constant $\edyn \sim 1\%$ matches the observations at least as good as any of the more sophisticated models that we tested. If one considers only low-pressure environments---such as the solar neighborhood---then the turbulence regulated models can provide a superior description of the observations with respect of a constant \edyn\ as long as the model's fudge factors (in particular the overall normalization) are appropriately adjusted  \citep[this requirement has also been pointed out by][]{Leroy17b, Ochsendorf17}. However, in the vein of \citet{Barnes17}, we have to conclude that the falsification of these star formation theories is currently obstructed by the lack of consensus on the values of their free parameters.

In line with several recent studies, our observations show that current models of turbulence regulated star formation---based on idealized assumptions of cloud-scale density structure and turbulence, and assuming a stationary star formation rate---do not fully capture observations across a diverse range of galactic and extragalactic star-forming environments \citep[][]{LeeE16, Barnes17, Leroy17b, Ochsendorf17}. The discrepancy may reflect observational limitations in constraining the relevant physical parameters (e.g., uncertain mass-to-light conversion and beam diluted measurements), however, it could also be of physical origin in that cloud-scale turbulent properties insufficiently reflect the dynamic ``boundary'' conditions of star-forming molecular clouds and that additional large-scale processes are relevant (e.g., implying environmental changes in the sonic and Alf\'{e}nic Mach number and the nature of turbulence driving). Moreover, it may imply the relevance of other parameters (e.g., the cloud lifetime or feedback efficiency) that have not yet been considered by analytic theories.

\section{Outlook}
\label{sec:outlook}

Within the next years, it will become possible to study the relationship between galactic structure, ISM properties, and star formation in unprecedented detail. With ALMA, we can now map the molecular gas (traced by CO emission) at cloud-scale resolution across entire galaxies (and samples thereof). Performing homogeneous analyses of such data with refined methodology \citep[e.g.,][]{Leroy16} can lead to accurate determinations of ISM properties such as the virial state of gas \citep{Sun18}. Comparison of these properties in smaller (kpc-scale) patches of galaxies allows a more accurate description of (their correlation with) the local galactic host environment. Additionally, observations with VLT/MUSE and JWST can trace early (massive) star formation, which enables studies of individual star-forming clouds as routinely performed in the Milky Way. Observations of higher critical density tracers (e.g., HCN) provide information on the gas density distribution. While cloud-scale mapping of dense gas tracers remains a challenge even with ALMA, coarser (kpc-scale) observations provide valuable insight too \citep[e.g.,][]{Usero15, Bigiel16, Gallagher18}, and can be further refined by modeling of the unresolved ISM structure \citep[][]{Leroy17a}. All these steps are goals of the PHANGS$^{\ref{footnote2}}$ collaboration.

In addition, observations that zoom in on the clouds in both \hi\ and CO tracers are desired. Especially for diffuse clouds dominated by ambient pressure, it is of interest to know their internal structure, look for gravitational bound cores, and search for \hi\ shielding envelopes around CO-bright cores. A first such assessment has been possible using parsec-scale CO observations with ALMA of star-forming regions in the Local Group dwarf galaxy NGC 6822 \citep{Schruba17}. Moreover, we want to know whether the CO is a continuous part of an \hi\ turbulent cascade or whether the chemical transition coincides with a transition in energy balance.

To further expand and test the model of turbulence regulated star formation, it is desirable that the model fudge-factors, which so far have been constrained with idealized, turbulent box simulations and vary significantly between different calibrations \citep[e.g.,][]{Federrath12}, are calibrated with galaxy-scale simulations. Moreover, we require predictions along the evolutionary tracks of star-forming clouds to interpret the scatter in cloud-scale resolved observations. Direct observational constraints on the evolutionary timeline of molecular clouds and the efficiency of stellar feedback are currently being derived \citep[e.g.,][]{Kruijssen14a, Kruijssen18, Kruijssen19b, Chevance19}. Additionally, obtaining Mach number measurements across a wide dynamical range of spatial scales (i.e., through the size-linewidth relation) will provide critical input for testing the role of turbulent energy driving and dissipation in turbulent star formation models. This will require high dynamic range ALMA observations. Finally, these different aspects need to be brought together to form a self-consistent theory that connects both galactic and cloud-scale ISM structure and star formation; with first progress underway \citep[e.g.,][and others]{Ostriker10, Semenov16, Krumholz18}.

\section{Summary}
\label{sec:summary}

We have built the largest compilation of measured molecular cloud properties and their galactic host environments, covering the Milky Way and seven nearby galaxies. Using mass-weighted mean molecular cloud properties for entire galaxies or distinct subregions therein, we study (a) the environmental dependence of the dynamical state of molecular clouds, and (b) the impact of the clouds' dynamical state on the global star formation activity, which we use to test analytic models of turbulence regulated star formation. Our main findings are:

\begin{itemize}
\item Molecular clouds are in ambient pressure-balanced virial equilibrium. In gas rich, molecular-dominated, high-pressure regions of galaxies, clouds are near virialization considering only their self-gravity ($\avir \approx 1-2$). Clouds in lower surface density, atomic gas dominated, low-pressure environments have $P_{\rm int} \sim P_{\rm ext}$ (resulting in $\avir \approx 3-20$) and are thus pressure confined.
\item The SFE per free-fall time is low ($\eff \approx 0.1\% - 1\%$), has significant ($1-2$~dex) scatter, and systematically varies with \avir\ and host galactic properties. For low-pressure, atomic-dominated regions we find a common trend (anti-correlation) between \eff\ and \avir.
\item Models of turbulence regulated star formation can provide a good match to the observations when considering only low-pressure, atomic-dominated regions and allowing for an \emph{ad hoc} adjustment in the overall normalization. The low observed \eff\ for clouds in high-pressure, molecular-dominated regions is not well reproduced by these models even when considering the maximal plausible range in the additional model parameters (Mach number, turbulence driving, and magnetic field strength) that could not be constrained by the available data. This suggests the importance of additional physical parameters not yet considered by these models.
\end{itemize}

We highlight that studies linking cloud-scale ISM properties, star formation, and their galactic environment are in their infancy. Within the next years, we expect many more insights from multi-wavelength, cloud-scale imaging surveys of nearby galaxies with ALMA, VLT/MUSE, and JWST. At the same time, the expansion of numerical simulations and analytic modeling to multi-scale models will be indispensible for interpreting the observations and building a self-consistent theory of galactic and cloud-scale ISM properties and star formation.

\acknowledgments

We thank the anonymous referee for a constructive report that led to improvements of the manuscript. JMDK gratefully acknowledges funding from the German Research Foundation (DFG) in the form of an Emmy Noether Research Group (grant number KR4801/1-1) and the Sachbeihilfe (grant number KR4801/2-1), from the European Research Council (ERC) under the European Union's Horizon 2020 research and innovation programme via the ERC Starting Grant MUSTANG (grant agreement number 714907), and from Sonderforschungsbereich SFB 881 ``The Milky Way System'' (subproject B2) of the DFG. The work of AKL is partially supported by the National Science Foundation under Grants No. 1615105, 1615109, and 1653300.




\bibliographystyle{aasjournal}


\end{document}